\documentclass[usenatbib,usegraphicx,useAMS]{mn2e}
\usepackage{amssymb,url,times}

\def\be{\begin{equation}}
\def\ee{\end{equation}}
\def\no{\noindent}
\def\ltsima{$\; \buildrel < \over \sim \;$}
\def\lsim{\lower.5ex\hbox{\ltsima}}
\def\gtsima{$\; \buildrel > \over \sim \;$}
\def\gsim{\lower.5ex\hbox{\gtsima}}

\def\apj{Ap.\ J.}
\def\apjl{Ap.\ J.\ Lett.}
\def\apjs{Ap.\ J.\ Suppl.}
\def\mnras{Mon.\ Not.\ Roy.\ Astro.\ Soc.}
\def\aap{Astron.\ Astrophys.}

\def\apss{Ap\&SS}

\title[tidal heating in white dwarf binaries]{Constraining white dwarf viscosity through tidal heating in detached binary systems}

\author[Dall'Osso \& Rossi]{Simone Dall'Osso$^{1,2}$\footnote{simone.dallosso@uni-tuebingen.de} \& Elena M. Rossi$^{3}$ \\
$^1$  Racah Institute of Physics, The Hebrew University of Jerusalem, Jerusalem 91904, Israel \\
$^2$ Theoretical Astrophysics, University of T\"{u}bingen, Auf der Morgenstelle 10,
D-72076 T\"{u}bingen, Germany\\
$^3$ Leiden Observatory, Leiden University, P.O. Box 9513, 2300 RA , Leiden, The Netherlands }
\date{Submitted: Revised:  Accepted:}
\pagerange{\pageref{firstpage}--\pageref{lastpage}} \pubyear{2013}

\begin{document}
\label{firstpage}
\bibliographystyle{mn2e}
\maketitle

\begin{abstract} 
Although the internal structure of white dwarfs is considered to be generally well understood, the source and entity of viscosity is still very uncertain.   
We propose here to study white dwarf viscous properties using short period ($ <$ 1 hr), detached white dwarf binaries, such as the newly discovered
$\sim 12.8$ min system (J0651). These binaries are wide enough that mass transfer has not yet started 
but close enough that the least massive component is subject to a measurable tidal deformation. 
The associated tidal torque transfers orbital energy, which is partially converted into heat by the action of viscosity  within the deformed star.
As a consequence, its outer non-degenerate layers expand, and the star puffs up. 
We self-consistently calculate 
the fractional change in radius, and the degree of asynchronism (ratio of stellar to orbital spin) as a function of the 
viscous time. Specializing our calculations to J0651, we find that the discrepancy between the measured radius of the secondary star and He white dwarf model predictions can be interpreted as tidal inflation if the viscous timescale is either $ \sim 2 \times 10^{5}$ yr or $\sim 10^{4}$ yr.
Such values point to a non-microscopic viscosity, possibly given by tidally induced turbulence, or by magnetic field stresses with a magnetic field strength of 10-100 Gauss.
Fortunately, these two timescales produce very different degree of asynchronism, with the shortest one, bringing the system much closer to synchronisation.
A measurement of the stellar spin can thus univocally determined the mean viscosity.
Currently, we may exclude a middle range viscous time of a few $10^4$ yr, which would give a radial inflation of $\sim 10\%$, which is not observed. 
Extrapolating the secondary's expansion we predict that the star will fill is Roche lobe at a separation  which is $\sim 1.2-1.3$  smaller than the current one.  Applying this method to a future sample of systems can allow us to learn whether viscosity changes with mass and/or nuclear composition.
\end{abstract}


\section{Introduction}
\label{sec:intro}
Short-period white dwarf binaries are very interesting systems. They are emitters of gravitational waves (GW), and candidate progenitors of
supernovae Ia (Webbink 1984; Iben \& Tutukov 1984) and .Ia (Bildsten et al. 2007). 
When their orbital period is of tens of minutes and mass transfer has not yet started, they become clean laboratories to study the reaction of their internal
structure to tidal forces. In turn, from this reaction a lot can be learnt on the internal properties of these stellar remnants.
Recently, the ``Extreme Low Mass (ELM)'' Survey  (Brown et al. 2010; Kilic et al. 2010) has increased to 24 the number of known systems
that will merge within a Hubble time (Kilic et al. 2012). In particular, there are currently three detached systems with orbital period $P$ less than 40 minutes: 
J0651 ($P \approx 12.75$ min, Brown et al. 2011), J1630 ($P\approx 39$ min, Kilic et al. 2011a) and J0106 ($P \approx 39$ min, Kilic et al. 2011b). Among these, 
the best case study is J0651. First, tidal elongation of the secondary\footnote{Here and thereafter, we refer 
to the least massive member of the binary as ``secondary". The most massive is therefore the ``primary" member.} was directly measured from the source lightcurve. 
Second, it is an eclipse system, which allows for a precise and model independent determination of the stellar radii  (see Table 1). 
Finally, the system has the shortest GW-driven merger timescale (a few Myr), which makes 
evolutionary processes particularly relevant.
% and, for this reason, modelling its evolution can be particularly sensitive to 
%the timescale of viscous processes, which is expected to be only slightly shorter.
%which start to be comparable with possible viscous timescales. FORSE LA TOGLIAMO PROPRIO questa ultima frase?}
%For this reason, its modelling is very sensitive to the value of the viscous timescale. 

We also find very intriguing that the measured radius of the secondary
$R_2= 0.0353 \pm 0.0004 R_{\odot}$ is $5\%$ larger than that predicted by the best matching He WD model ($\approx 0.0337 R_{\sun}$, Panei et al. 2007). This discrepancy is 
significant at the 4$\sigma$ level. We will show that it can be explained as due to tidal heating with plausible values of the viscous timescale.

In white dwarfs, possible values of the dynamical viscosity span a wide range, and observations have not yet been able to constrain it.
Values of $\mu=(10^{11}-10^{14})$ g cm$^{-1}$ s$^{-1}$ have been suggested (e.g. Iben, Tutukov, \& Fedorova 1998), many orders of magnitude larger than classical radiative 
or plasma viscosity (Kopal 1968; Durisen 1973), which would imply the possible role of turbulence (Zahn 1966, 1977), shear (Press, Wiita \& Smarr 1975) or magnetic 
(B $\sim$ Gauss, Sutantyo 1974) viscosity, and would correspond to a viscous timescale $t_{\rm v} \equiv R_2^2 \bar{\rho}/\mu \sim$ ten up to ten thousands of years, depending 
on the white dwarf radius and average density $\bar{\rho}$.
Values as large as $\mu \approx 10^{18}$ g cm$^{-1}$ s$^{-1}$ have even been proposed (t$_{\rm v} \sim 1$ day), which would require magnetic field strength of $10^3-10^6$ Gauss 
(Smarr \& Blandford 1976). Given these timescales, viscosity is expected to play a major dynamical role for the observed detached white dwarf binaries, and therefore we can study its effect using these systems. We will focus in particular on J0651, for which tidal effects are strongest and the rapid orbital evolution makes it particularly sensitive to viscous times $\lesssim$ a few Myr (or, say, $\mu \gsim 10^9$ g cm$^{-1}$ s$^{-1}$). 

In our first paper, Dall'Osso \& Rossi (2013) (hereafter Paper I), we studied tidal interactions in merging compact object binaries, with a main focus on neutron star secondaries.
For  such systems, when the orbital period is less than $\sim10$ s, the viscous timescale ($t_{\rm v} \sim 1$ yr) is longer than the GW-driven inspiral time. 
The latter is $\sim 10^{-2}$ s at merger, hence the neutron star can be considered inviscid when studying the tidal interaction in this final phase.
In this paper, we extend our previous model to account for the effect of viscosity, enabling us to trace the evolution of a binary system at any separation, either when the 
viscous timescale is longer or shorter than the inspiral time.

The paper is organized as follows.
In \S 2, we reproduce the expected tidal deformation in J1651.
This comparison is a sanity check for our model, since it should be independent of unknown properties like the stellar viscosity.
In \S 3, we model the system dynamics when the tidal torque is both induced by viscosity and gravitational wave orbital decay.
In \S 4, we  predict the effects of viscosity, both the tidal heating and the spin up of the secondary.  
We show how data comparison allow us to constrain the viscous timescale.
Finally, we draw our conclusions in \S 5.

\section{Tidal interaction}
\label{sec:tidal_deformation}

Let us consider a binary white dwarf system, at a separation where mass transfer has not yet started. 
For a system like J0651, we will show that this implies periods larger than $\sim 10$ min.
The primary and the secondary have masses and radii $M_1$ and $M_2$ and $R_1$ and $R_2$ respectively. We model the stars as polytropic structures
with index $n=3/2$, appropriate for low mass (i.e. non-relativistic) white dwarfs. The global deformability of the
secondary is measured by its Love number,  $\kappa_2 \approx 0.15 $, which depends on the star density profile (Love 1909, Verbunt \& Hut 1983, Hinderer 2008).

In this section, we report the calculation presented in Paper I of the evolution of the tidal deformation (referred to as ``bulge'') of the secondary star
as a function of the orbital separation. We work in the limit of small deformations, 
$\eta= h/R_2 \ll 1$,
where $h$ is the ``height'' of the tidal bulge, i.e. the elongation with respect to the unperturbed radius. The fractional height $\eta$
is a function of the orbital separation, $a$, and can be linked to the semi-major axis of the white dwarf by
$$R(a) = R_2 [1+\eta(a)]. $$
There exists a maximum $\eta$ beyond which the star's self gravity is not able anymore to counteract the tidal pull, and the star falls apart. 
This happens at the so called ``tidal radius'' $a_{\rm T}$, defined as $a_{\rm T} \approx 2.15 R_2 \,q^{-1/3}$,
where q = $M_2/M_1$.  The corresponding value of $\eta_{\rm T} \approx$ 0.42. For J0651 this separation is only $\simeq 2.8 \times R_2$, and mass transfer
starts well before the system can reach tidal disruption (see Table 1).

We sketch here the derivation of $\eta(a)$ as a function of the orbital separation. 
The secondary's tidal bulge is a result of the work done by 
the first term in the expansion of the primary's 
newtonian potential $\psi_{\rm T}$. If we consider a two dimensional problem, in the orbital plane of the binary,
\be
\psi_{\rm T} = \frac{G M_1}{a} \left(\frac{r}{a}\right)^{2} \frac{3\cos^2(\theta)-1}{2}.
\label{eq:psit}
\ee

\no
(Alexander 1973 and references therein), 
where the polar coordinates (r, $\theta$) are centered in the centre of mass of the secondary, and
$\theta$ is measured from the semi-major axis of the bulge. In eq. \ref{eq:psit}, $0 \le r \le R_2$.

This tidal pull causes a decrease in the binding energy of the originally unperturbed 
secondary. In turn, the change in the secondary's external potential, $\psi_*$,

\be
\psi_{*} = \kappa_2 \left(\frac{R_2}{r}\right)^3 \times \psi_{\rm T},
\label{eq:psistar}
\ee
is a result of the tidal bulge: since the two stars are now closer to each other, the orbital binding energy increases, 
further drawing from the stellar binding energy.

\begin{figure}
\includegraphics[width=\columnwidth]{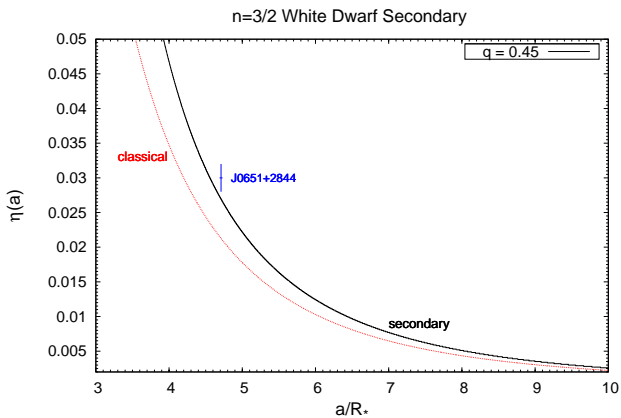}
\caption{Tidal deformation  as a function of the orbital frequency, for the secondary WD. The black line is 
our calculation (eq.\ref{eq:define-eta}), while the red line
is the classical approximation (eq.\ref{eq:classical_eta}). The system parameters are those of J0651 (see Table 1) and 
its observed tidal deformation is marked in blue, where we also 
indicated the error bars at one $\sigma$. The chosen adiabatic index is $n=3/2$.}
\label{fig:eta}
\end{figure}

 In the system's energy budget,
the additional energy terms $\psi_{\rm T}$ and $\psi_*$ are thus balanced by
the decrease in the stellar binding energies.   
This statement of energy conservation reads,
\begin{equation}
\label{eq:tidal-energy-balance}
M_2 \psi_{\rm T} + M_1 \psi_* =  \frac{\gamma GM_2^2}{R_2} - \frac{\gamma GM^2_2}{R},
\end{equation}
where $\psi_{\rm T}$ and $\psi_*$ are evaluated along the line of centres ($\theta =0$),
with $r=a$  and stellar radius $R$ (instead of $R_2$) in  eq. \ref{eq:psistar}, and with $r=R$ in eq. \ref{eq:psit}.
The structural constant in the self gravity term (right-hand site)
is $\gamma=3/(5-n) =6/7$ (cfr. Lai \& Shapiro 1995).
A manipulation of eq. \ref{eq:tidal-energy-balance}
leads to the following algebraic equation for $\eta$,
\begin{equation}
\label{eq:define-eta}
\kappa_2 (1+\eta)^6 + q \left(\frac{a}{R_2}\right)^3 (1+\eta)^3 - q^2 \gamma
\left(\frac{a}{R_2}\right)^6 \eta = 0 \, ,
\end{equation}
For a given polytropic index $n$ and the mass ratio $q$, we can solve
numerically the above equation to obtain $\eta(a)$. In Fig.\ref{fig:eta} we show 
our results for the the detached WD-WD system J0651, whose observed parameters are listed in Tab. \ref{tab:1}.
The agreement of the observed $\eta \approx 3 \%$ with our prediction (black line) is quite satisfactory. Instead, the classical approximation of $\eta$,
\begin{equation}
\frac{h}{R_2} =  \frac{\psi_{\rm  T}}{(\beta GM_2/R_2)},
\label{eq:classical_eta}
\end{equation}
where $\phi_*$ is not included, fails to describe the final stages of the
inspiral (red line).

\begin{table}
\begin{center}
\begin{tabular}{l l}
\hline \hline
  J0651 physical parameters \\ \hline \hline
  $M_2 =$ $0.25 M_{\sun}$ &  $P =$  $765.206(3)$ s \\  
  $M_1 =$ $0.55 M_{\sun}$ & $\omega_{\rm o} =$  $0.00821110(3) $~ rad s$^{-1}$\\  
  $q =M_2/M_1$ $= 0.45$ & $a \approx 4.7 R_2$ \\  
  $R_2 =$  $0.0353(4) R_{\sun} $  & $ a_{\rm Roche} \approx 3.76  R_2$  if t$_{\rm v} = 170$ kyrs \\  
 $\eta \approx 0.033  $   & $ a_{\rm Roche} \approx 3.96  R_2$ if t$_{\rm v} = 11$ kyrs \\
  \hline  
\end{tabular}
\end{center}
\caption{The detached WD system J0651. List of the measured parameters, which we use in this paper (Brown et al. 2011). The orbital separations at which (depending on $t_{\rm v}$) the secondary will fill its Roche lobe, $a_{\rm Roche}$, are theoretically derived in Sec. 4.3. The numbers in parenthesis indicate 1$\sigma$ errors on the last digit. The current orbital separation, $a$ and $a_{\rm Roche}$ are expressed in units of R$_2$.}
\label{tab:1}
\end{table}

%%%%%%%%%%%%%%%%%%%%%%%%%%%%%%%%%%%%%%%%%%%%%%%
%%%%%%%%%%%%%%%%%%%%%%%%%%%%%%%%%%%%%%%%%%%%%%%%%%%%%%%%%%%%%
%%%%%%%%%%%%%%%%%%%%%%%%%%%%%%%%%%%%%%%%%%%%%%%%%%%%%%%%%%%%%%%%%%%%%%%%%%%%%%%%%%%%%%%%%%%%%%%%%%%%%%%%%%%%%%%%%%%%%%%%%%%%%%%%%%%%%%%%%%%%%5
\section{The system dynamics}
\label{sec:dynamics}
In this section, we describe the dynamics of a detached binary system. 
The equations we derive here will allow us to calculate the tidal dissipation and the consequent
thermal expansion of the mean radius of the secondary.

The dynamics of an {\em inviscid} system at coalescence is described in Paper I. 
In that case, the orbital shrinkage is mainly due to emission of GWs. The rapid orbital decay causes a continuous speed -up of 
the orbital frequency with the bulge constantly lagging behind the line of the centres. This (small but finite) misalignment of 
the bulge axis with the line of the centres induces a continuous tidal torque which speeds up the bulge. At equilibrium the bulge tracks the acceleration of the orbit, hence $\dot{\omega}_{\rm b}  = \dot{\omega}_{\rm o}$. We called it a {\em GW-induced tidal torque}.

In white dwarf binaries 
with orbital period $ > 10$ minutes, the viscous timescale $t_{\rm v}$ is generally shorter than the timescale for GW-driven orbital evolution, $t_{\rm GW} \equiv 
\omega_{\rm o}/\dot{\omega}_{\rm o}$
\be
\label{taugw}
t_{\rm GW} \approx  4 \times 10^{5}~{\rm yr}~\left[\frac{\omega_{\rm o}}{0.00821~{\rm rad/s}}\right]^{-8/3} \frac{(1+q)^{1/3}q^{2/3}} {m^{5/3}_2},
\ee
(Peters 1964), where $m_2 = M_2/M_{\sun}$ and we specialised to a binary with a period of $P=12.8$ min. Hence, far from being negligible, viscosity is expected to largely determine the tidal interaction in these systems. Interestingly,
J0651 may be in an intermediate regime: its inspiral timescale,  $t_{\rm GW} \approx 3$ Myr, is sufficiently short that tidal coupling might receive a significant contribution also from the GW-induced tidal torque. This makes J0651 an excellent case to study tidal interactions in detail, and to constrain WD viscosity.

We generalise here the dynamics described in our previous paper, including the action 
of the viscous torque $N_{\rm v}$, which drains angular momentum from the bulge and transfers it to
the star's rigid rotation (i.e. it increases its spin).

The net torque acting on the bulge, and changing its angular momentum $J_{\rm b}$,  is
\be
\frac{d J_{\rm b}}{d t} = N_{\rm T} - N_{\rm v},
\label{eq:djbdt}
\ee
where the {\em tidal} torque, $N_{\rm T}$, speeds up the bulge \footnote{By transferring to it angular momentum extracted from the orbit.}, 
and the viscous torque $N_{\rm v}$ resists the motion and tends to slow it down. 
The bulge moment of inertia  $I_{\rm b}$ is, for small $\beta$ and small tidal deformation $\eta$,
$$ I_{\rm b} = \frac{6 \pi^2 \kappa_2}{(1+q)^2} \left(\frac{\omega_{\rm b}}{\omega_*}\right)^4 M_2 R^2_2,$$
(Paper I eq.4). The change of angular momentum ($J_{\rm b} = I_{\rm b} \omega_{\rm b}$) 
can therefore be expressed as $dJ_{\rm b} /dt = 5 I_{\rm b} \dot{ \omega}_{\rm b}$.

Let's now specify the two terms on the right hand side of eq. \ref{eq:djbdt}.
The tidal torque arises when the bulge angular speed $\omega_{\rm b} \ne \omega_{\rm o}$, 
which causes a misalignment by an angle $\beta$ between the bulge axis and the line of the centres. It can be derived from
$N_{\rm T}= \nabla_{\theta}\psi_{\rm *}(a) \times \mathbf{a}$, which gives the following expression
\begin{equation}
N_{\rm T}  \simeq N \beta \left(\frac{\omega_{\rm o}}{\omega_*}\right)^4,
\label{eq:Ntidal}
\end{equation}
where the numerical constant $N = {3 \kappa_2 \omega_*^2 M_2
  R^2_2}/{(1+q)^2}$, and we use the property that for our WD systems $\beta \ll 1$ (see Paper  I, eq. 3).
The {\em effective} lag angle $\beta$ in the above expression should be regarded
as a global property  of the star like, e.g., its total deformability $\kappa_2$. 
Finally, the angular momentum which is subtracted from the bulge by $N_{\rm v}$ is transferred to the spin, 
\be
N_{\rm v} = \frac{dJ_{\rm s}}{dt} \equiv \frac{J_{\rm b}}{t_{\rm v}},
\label{eq:spin}
\ee
where with the last step we define the viscous timescale. The WD spin is $J_{\rm s} = I_* \omega_{\rm s}$, where the unperturbed moment of inertia is 
$I_* \approx 0.2 M_2 R^2_2$ and $\omega_{\rm s}$ is the spin angular frequency. 

The ultimate source of angular momentum and energy in tidal interactions is the binary orbit. Its angular momentum $J_{\rm o}$ 
decreases because of the existence of a tidal torque and because it can be transported away by GW emission,
\be
\frac{d J_{\rm o}}{dt} = - \left(N_{\rm T}+ N_{\rm GW}\right) , 
\label{eq:dJtot}
\ee
where $J_{\rm o} = I_{\rm o} \omega_{\rm o}$,
\be
N_{\rm GW}= \frac{1}{3} I_{\rm o} \frac{\omega_{\rm o}}{t_{\rm GW}},
\label{eq:Ngw}
\ee
and  the orbital moment of inertia is $$I_{\rm o} = M_2 M_1^{2/3} G^{2/3} (1+q)^{-1/3} \omega_{\rm o}^{-4/3}.$$ 
At equilibrium the bulge speeds up at the same rate as the orbit shrinks, $\dot{\omega}_{\rm b} = \dot{\omega}_{\rm o}$. 
Using this condition and eqs. \ref{eq:djbdt} and \ref{eq:dJtot}, it is easy to derive that in absence of GW emission ($N_{\rm GW}=0$) 
the tidal torque, due only to the WD viscosity, is
\be
N_{\rm T} = \frac{N_{\rm v}}{1-x} \equiv N_{\rm T,v} ~,
\label{eq:Ntv}
\ee
where $$x = \frac{15 I_{\rm b}}{I_{\rm o}} = \frac{90 \pi^2 \kappa_2 R^8_2 \;\omega_{\rm o}^{16/3}}{(1+q)^{5/3} G^{8/3} M_2^2 M_1^{2/3}}.$$ 
When, in addition, a dynamical component of the tidal torque arises due to the emission of GWs ($N_{\rm T,GW}$), the total tidal torque can be expressed as 
\be
N_{\rm T} = N_{\rm T,v} + N_{\rm T,GW} = N_{\rm T,v} \left(1 + \frac{ 5 t_{\rm v}}{t_{\rm GW}}\right) = N_{\rm T,v} \left(1 + \tau \right),
\ee
where the last step defines  the parameter $\tau = 5 t_{\rm v}/t_{\rm GW}$, measuring the relative contribution of N$_{\rm T,GW}$ to the total tidal torque.
In J0651 for example, we obtain $\tau \approx 0.2$ for t$_{\rm v} = 10^5$ yrs, scaling linearly with the poorly constrained value of the viscous time. 
%depending on th{{\e poorly constrained value of $t_{\rm v}$, in therange $10^4$ yr $<t_{\rm v}<10^6$ yr. 
Thus, the tidal torque would be dominated by viscous effects as long as t$_{\rm v} \ll 10^5$ yrs, say, while for larger values of t$_{\rm v}$
the GW-induced torque becomes progressively important, and even dominant at t$_{\rm v} \gsim 5\times 10^5$ yrs.
For wider systems, on the other hand, N$_{\rm T,GW}$ is generally expected to be unimportant, due to the steep increase of t$_{\rm GW}$ with the orbital period.
 
Finally, we can evaluate the relative importance of the tidal torque versus GW emission for causing the orbital decay,

\be
\frac{N_{\rm GW}}{N_{\rm T}} =  \frac{\tau}{x} \frac{1}{(1+\tau) (1-\alpha)} \, ,
\ee
where $\alpha = \omega_{\rm s} / \omega_{\rm o}$ measures the asynchronism of the secondary.
For low mass systems with orbital periods longer than $\sim 10$ minutes, $x \ll 1$ and $\tau \lesssim 1$.
For J0651 we get $x \approx 4 \times 10^{-3}$, thus, as long as t$_{\rm v} > 2 \times 10^3$ yrs, the effect of the tidal torque on the 
orbital evolution is at best marginal, even if $\alpha = 0$. However, as we will see below, the asynchronism is expected to be very close to unity if t$_{\rm v} \leq 10^4$ yrs. Hence, GW emission is always dominant in determining the measured orbital shrinkage of J0651.
%this effect becomes rapidly important for values of t$_{\rm v} < 10^4$ yr, 
%and even dominant when t$_{\rm v} < 2 \times 10^3$ yr (or $\mu > 10^{12}$ g cm$^{-1}$ s
%$^{-1}$). For such large viscosities, even the orbital evolution of J0651 would be mostly %driven by the action of the tidal torque, making GW emission a secondary effect.

\section{The effects of viscosity}

The WD internal viscosity plays an all-important role in determining its tidal evolution. Viscous dissipation of the bulge energy causes a progressive
synchronisation of the WD spin with the orbital period. The released heat in turn affects the internal energy balance and can even induce measurable changes of 
the WD global properties, such as its outer radius or its luminosity. 

\subsection{Degree of asynchronism}

As we discussed in detail in sec. \ref{sec:dynamics}, the internal viscous torque changes the spin of the star by transferring to it 
angular momentum, drained from the tidal bulge. The temporal evolution of the asynchronism $\alpha$ is

\be
\frac{\dot{\alpha}}{\alpha} = \frac{\dot{\omega}_{\rm s}}{\omega_{\rm s}} -  \frac{\dot{\omega}_{\rm o}}{\omega_{\rm o}}, 
\ee
which requires to simultaneously track the spin and orbital evolution.
The former is governed by eq. \ref{eq:spin}, and depends only on the viscous torque,
\be
\frac{\dot{\omega}_{\rm s}}{\omega_{\rm s}} = \frac{N_{\rm T,v}} {I_*} \frac{(1-x)} {\omega_{\rm s}},
\label{eq:dots}
\ee
where we used eq. \ref{eq:Ntv} and the definition
$dJ_{\rm s}/dt = I_* \dot{\omega}_{\rm s}$.
The orbital angular frequency, instead, can change both under the action of GW emission and of the tidal torque (eq. \ref{eq:dJtot}). 
Since, $dJ_{\rm o}/dt = -1/3 I_{\rm o} \dot{\omega}_{\rm o}$, we can derive that,

\be
\frac{\dot{\omega}_{\rm o}}{\omega_{\rm o}} = \frac{3 N_{\rm GW}}{I_{\rm o} \omega_{\rm o}} \left(1+ \frac{{\rm N}_{\rm T}}{{\rm N}_{\rm GW}}\right)= 
\frac{\left[1+ (1-\alpha)\, (1-\tau)\,x/\tau\right]}{t_{\rm GW}}.
\label{eq:doto}
\ee
Finally, from eqs. \ref{eq:dots} and ~\ref{eq:doto} and after some manipulation  we obtain the temporal evolution of $\alpha$,
\be
\dot{\alpha} =\frac{(1-\alpha)}{t_{\rm v}} Q  (1-x) \omega_{\rm o}^4 - \frac{\alpha}{t_{\rm GW}}- \frac{\alpha (1-\alpha)}{{\rm t}_{\rm v}} \frac{x (1+\tau)}{5} ,
%(1-x)^{-1} \left(1+\frac{x}{\tau}\right),
\label{eq:dot_alpha}
\ee
where $Q = 30 \kappa_{\rm 2} \pi^2/[(1+q)^2 \omega_{*}^4]$.

%%%%%%%%%%%%%%%%%%%%%%%%%%%%%%%%%%%%%%%%%%%%%%%
\begin{figure}
\includegraphics[width=\columnwidth]{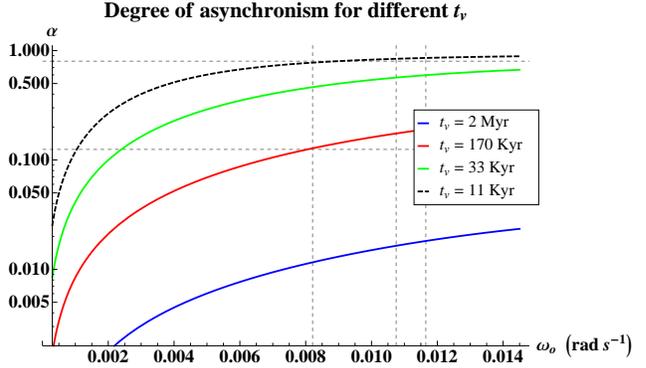}
\caption{Degree of asynchronism $\alpha = \omega_{\rm s}/\omega_{\rm o}$, 
as a function of the orbital frequency, for four values of the viscous timescales (see legend).
The system parameters are the same as Fig.\ref{fig:eta} (those for J0651, see Table 1). The leftmost vertical dashed line indicates  the current orbital frequency of J0651. The two dashed vertical lines to the right indicate the orbital frequencies at which mass transfer will start ($P \sim 9.7$ min and $\sim 9$ min) for two different values of the viscous time, t$_{\rm v}  \approx 170$ kyrs and $\approx$ 11 kyrs respectively (see below, $\S$ \ref{sec:inflation}). The two horizontal dashed lines mark the asynchronism predicted for J0651 at its observed separation, if the viscous timescale has either of the two values. $\alpha \approx 0.8$ corresponds  to the shorter t$_{\rm v}$,
$\alpha \approx 0.125$ to the longer one.}
%%$t_{\rm v} \approx 2 \times 10^{5}$ yr (lower value) or $t_{\rm v} \approx 10^{4}$ yr (upper value). See also %section \ref{sec:inflation}.}
\label{fig:alpha}
\end{figure}
%%%%%%%%%%%%%%%%%%%%%%%%%%%%%%%%%%%%%%%%%%%%%%%%%%%%%%%%%%%%%
In general, as long as $t_{\rm GW} \gg t_{\rm v}$, the asynchronism decreases  ($\dot{\alpha}>0$) and $\alpha$ reaches 
the asymptotic solution of eq. \ref{eq:dot_alpha} at a given orbit, before the system changes its separation. 
On the other hand, for  $t_{\rm GW} \ll t_{\rm v}$ the viscosity is too slow to be able to efficiently 
transfer angular momentum to the star spin before the orbital separation changes. As a consequence, the asynchronism progressively 
increases ($\dot{\alpha} <0$). This latter case is relevant, for instance, for double neutron stars at coalescence (Paper I), 
which never reach tidal locking (e.g. Bildsten \& Cutler 1992). For systems considered in this paper, instead, $\alpha$ 
changes on a viscous timescale and it is possible to describe its evolution  as a function of $\omega_{\rm o}$, 
as a series of equilibrium solutions given by eq. \ref{eq:dot_alpha},
\be
\frac{d\alpha}{d\omega_{\rm o}} = \frac{\dot{\alpha}}{\dot{\omega}_{\rm o}} =  \, \frac{Q~ t_{\rm GW} (1-x) (1-\alpha)}{t_{\rm v} \left[1+(1+\tau)(1-\alpha) x/\tau\right]}\,\omega_{\rm o}^{3} -\frac{\alpha}{\omega_{\rm o}},
\label{eq:alpha_wo}
\ee
where the time evolution of $\omega_{\rm o}$ is given by eq. \ref{eq:doto}.
The solution is shown in Fig. \ref{fig:alpha}, 
for four values of the viscous timescale, $t_{\rm v}$ = 11 kyrs, 33 kyrs, 170 kyrs and 2 Myrs.
We start with $\omega_{\rm s} \approx 0$ at large separation, motivated by the long, $\sim$ hrs or more, spin periods of isolated WD dwarfs (Spruit 1998 and references
therein).
This figure shows that, only for very short viscous timescale $t_{\rm v} < 10^{4}$ yr, tidal synchronisation is (nearly) reached before the two stars come into contact (for J0651 the latter happens at $P\simeq 9$ min, if t$_{\rm v} \approx 11$ kyrs, or at $P\simeq 9.7$ min, if t$_{\rm v} \approx 170$ kyrs  , see $\S$ \ref{sec:inflation}) 
%{\bf NELLE CAPTIONS SCRIVIAMO 9.3 INOLTRE IN  5 LA RIGA NON INTERSECA $RL=DR$, INOLTRE CE NE DOVREBBERO ESSER DUE DI LINEE CORRISPONDENTI AI DUE TEMPI VISCOSI, O ALMENO DOBBIAMO CITARLI ENTRAMBI}). 
The leftmost vertical line marks the current orbital frequency of J0651: its degree of synchronisation is very sensitive to the value of the viscous timescale, going from $< 1\%$ if t$_{\rm v} > 10^6$ yrs, to $\gsim 80$\% if t$_{\rm v} \lesssim 10^{4}$ yr range.
\subsection{Internal heating}
\label{sec:heating}
The viscous torque also drains energy from the tidal bulge along with angular momentum. While part of this energy is transferred
to the WD spin another part is released locally, heating the inner layers of the star. The rate of heat release will thus be given by
the difference between the rate of energy extraction from the bulge and the rate of energy transfer to the spin. 
\be
W = N_{\rm v} \omega_{\rm b} - N_{\rm v} \omega_{\rm s} \approx N_{\rm T, v} (1-x) \,\omega_{\rm o} (1-\alpha) = \frac{c_2}{t_{\rm v}} ~\omega^6_{\rm o} (1-\alpha)^2
\label{eq:W}
\ee
where $c_2 = 6 \pi^2 \kappa_2 (1-x) M_2 R^2_2 / \left[\omega_* (1+q)\right]^2$  depends only on stellar parameters 
%and can be derived from $N$ (see sec. \ref{sec:dynamics}) 
and we used the expression for the lag angle (cf. Paper I), $$\beta= 2 \pi^2 \frac{\omega_{\rm o}}{({\rm t}_{\rm v}\omega^2_*)} (1-\alpha).$$

%%%%%%%%%%%%%%%%%%%%%%%%%%%%%%%%%%%%%%%%%%%%%%%
\begin{figure}
\includegraphics[width=\columnwidth]{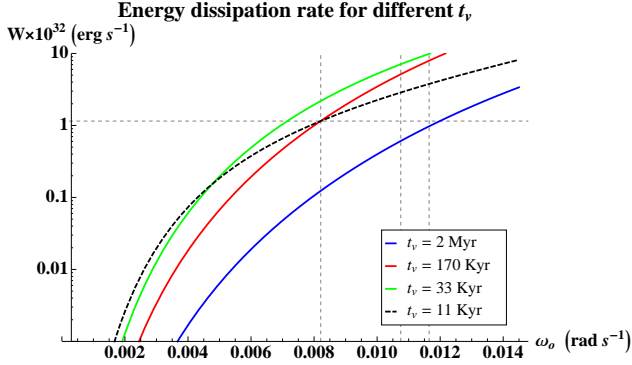}
\caption{Energy disspation rate (eq.\ref{eq:W})
as a function of the orbital frequency, for four values of the viscous timescales (see legend).
The system parameters are the same as in Fig.\ref{fig:alpha}. At the separation of J0651 (left dashed line), the energy dissipation is
$1.15 \times 10^{32}$ erg s$^{-1}$, or equivalently $\simeq 38\%$ of the WD luminosity. The vertical dashed lines have the same meaning as in the previous figure. }
%mark the orbital frequency at whichmass transfer will start ($P\simeq 9.3$ min).}
\label{fig:W}
\end{figure}
%%%%%%%%%%%%%%%%%%%%%%%%%%%%%%%%%%%%%%%%%%%%%%%%%%%%%%%%%%%%%

\begin{figure}
\includegraphics[width=\columnwidth]{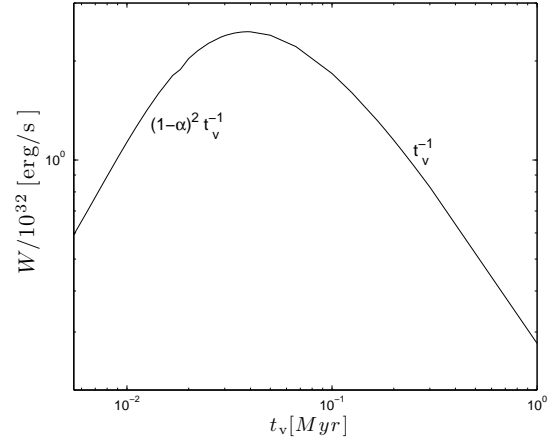}
\caption{The heating rate due to viscosity as a function of the viscous time in Myr, for J0651.
For long ($> 4 \times 10^{4}$ yr) viscous timescales the energy dissipation rate decreases with $t_{\rm v}$ as
$\propto t_{\rm v}^{-1}$. For shorter ones, the energy dissipation rate {\em increases} with $t_{\rm v}$ as $\propto (1-\alpha)^{2}/t_{\rm v}$. The peak is around 
$t_{\rm v} \approx 3.3 \times 10^{4}$ yr.}
\label{fig:W-tv}
\end{figure}

Once the evolution of the asynchronism is determined by eq. \ref{eq:alpha_wo}, the associated energy dissipation rate can be obtained straightforwardly. The 
resulting evolution of $W$ with the orbital period 
is plotted in Fig. \ref{fig:W} for the same four values of the viscous timescale used before (see figure caption for details). 
Viscosity has a double role in determining the energy dissipation rate, at a given $\omega_{\rm o}$. The viscous timescale appears explicitly in eq. \ref{eq:W}, 
through the term N$_{\rm v}$ (see eq. \ref{eq:spin}), and implicitly, through the term $(1- \alpha)$. 
While decreasing the viscous time would increase the coefficient in eq. \ref{eq:W}, it would also cause a smaller
value of $(1- \alpha)$, because spin synchronisation would be more efficient. 
The value of $W$ will be determined by a trade-off between these two competing effects.
To illustrate this, we show in Fig. \ref{fig:W-tv} the dependence of $W$ on the viscous timescale, with labels on different branches of the curve,
indicating the terms that determine the dependence on viscosity. As expected from the previous discussion, $W$ displays a maximum around a specific value of t$_{\rm v}$ $\sim 35$ kyr, and decays at both ends of the considered range. In particular, we obtain $W_{\rm max} \sim  2.5 \times 10^{32}$ erg s $^{-1}$ for t$_{\rm v} \sim 35$ kyrs, corresponding to $(dR/R)_{\rm max} \sim 0.095$.

%%%%%%%%%%%%%%%%%%%%%%%%%%%%%%%%%%%%%%%%%%%%%%%
%%%%%%%%%%%%%%%%%%%%%%%%%%%%%%%%%%%%%%%%%%%%%%%%%%%%%%%%%%%%%

\subsection{Inflation of the secondary white dwarf}
\label{sec:inflation}
The viscously dissipated energy ultimately inflates the outer --- non degenerate --- layers of the white dwarf. These are in radiative and hydrostatic equilibrium, which combined with the equation of state for ideal gases gives the following relation between the density $\rho$ and the temperature $T$,

 \be
 \rho \approx  1.52 \times 10^{-21} \left(\frac{M_2}{L} \right)^{1/2} T^{3.25}~{\rm g~cm^{-3}},
 \label{eq:a}
  \ee
  where $L$ is the WD luminosity, and here and in the following we adopt a WD composition with 90\% He and 10\% metals.  
  The deep interior of the star is instead electron degenerate and the transition between the two regimes occurs at the layer where 
  density and temperature are related by
 \be
 \rho_* \approx  4.8 \times 10^{-8} T_*^{1.5}~{\rm g cm^{-3}},
 \label{eq:b}
 \ee
 (see Shapiro \& Teukolski 1983).  
The non-degenerate layer is quite thin, and it can be shown that the transition radius, $R_{\rm t}$, and the stellar radius $R_2$ are related by
 
 \be
 R_2 = \frac{A R_{\rm t}}{R_{\rm t} T_* -A},
  \label{eq:c}
 \ee
 where $A \approx 2.64 \times 10^{-16} M_2/R_2$. Combining eq. \ref{eq:b} and eq. \ref{eq:c}, we can express the transition temperature in terms of the WD luminosity,
 $L \propto T_*^{3.5}$. Therefore, a perturbation $W$ to the  WD luminosity causes a fractional upward change in the transition temperature given 
by $W/L = 3.5 \times dT_*/T_*$.
 
The tidally-induced increase of the WD internal luminosity will affect the radiative and hydrostatic equilibrium of the non-degenerate envelope, inducing a change in
its radial extension, hence in the WD radius $R_2$. 
 The variation of the external radius as a consequence of a tidal dissipation rate $W$ can be calculated from eq. \ref{eq:c},
 \be
 dR_2 = \left[ \left(\frac{\partial R_2}{\partial T_*}\right)_{R_{\rm t} } + \left(\frac{\partial R_2}{\partial R_{\rm t}}\right)_{T_*}  \frac{dR_{\rm t}}{dT_*}\right] dT_*,
  \label{eq:dR_*}
 \ee
 from where  it is straightforward to derive 
 $$  \left(\frac{\partial R_2}{\partial T_*}\right)_{R_{\rm t} } = \frac{R^2_2}{A}, ~~~~ \left(\frac{\partial R_2}{\partial R_{\rm t}}\right)_{T_*} = \left(\frac{R_2}{R_{\rm t}}\right)^2.$$
 The relation between $T_*$ and $R_{\rm t}$ needs instead knowledge of the density radial profile in the envelope, which can be derived from the radiative 
transfer equation together with eq. \ref{eq:b}
 \be
 \frac{d\rho}{dr} = -3.25 \frac{A}{T} \frac{\rho}{r^{2}}.
 \ee
 Therefore,
 
 \be
  \frac{dR_{\rm t}}{dT_*} = \left( \frac{d\rho}{dr}\right)^{-1} \frac{d\rho_*}{dT_*} = - \frac{3}{6.5} \frac{R_{\rm t}^2}{A},
 \ee
where we used eq. \ref{eq:b} to evaluate  $d\rho_*/dT_*$. We can finally insert all pieces in eq. \ref{eq:dR_*} and 
derive the fractional increase of the stellar radius,

\be
\frac{dR_2}{R_2} = \frac{7}{13} \frac{R_2 T_*}{A} \frac{dT_*}{T_*}  \approx 1.9 \frac{R_2 T_*}{A} \frac{W}{L}.
\label{eq:dR2suR2}
\ee

\begin{figure}
\includegraphics[width=\columnwidth]{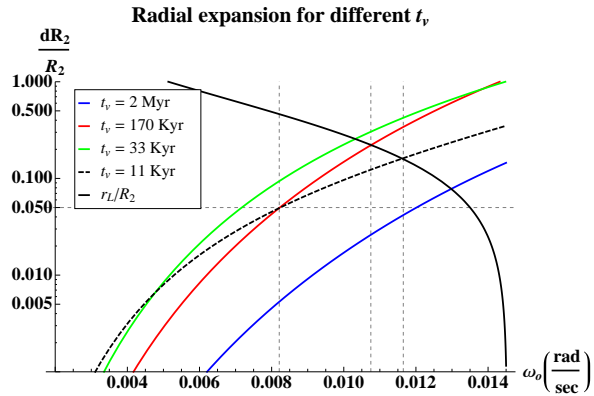}
\caption{ Fractional change of the outer radius of the WD (eq.\ref{eq:dR2suR2}) 
as a function of the orbital frequency, for different viscous times (see legend).
The solid black line is the fractional size of the Roche Lobe, $(r_{\rm L}-R_2)/R_2$.
The system parameters are the same as in Fig. \ref{fig:alpha}.
 The possible observed value for J0651 $dR_2/R_2 \simeq 5\%$ is marked with a dashed horizonatal line.  The orbital periods at which the Roche lobe is filled are readily found by the intersection of the solid black line with either the solid red curve or the black dashed curve. The  vertical dashed lines have the same meaning as in the previous figures.
}
\label{fig:drsur}
\end{figure}

The set of equations which comprises  eq. \ref{eq:dot_alpha}, eq. \ref{eq:alpha_wo}, eq. \ref{eq:W} and eq. \ref{eq:dR2suR2} 
allow us to self consistently track the viscous inflation of the mean stellar radius as a function of the orbital period.
In Fig. \ref{fig:drsur}, we show the radial inflation of the star as a function of the system separation for increasingly large viscous times. 

\begin{figure}
\includegraphics[width=\columnwidth]{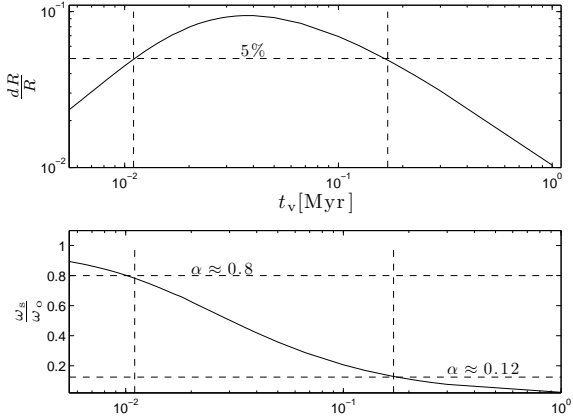}
\caption{ {\em Upper panel:} fractional change of the outer radius of the secondary in J0651, as a function of the viscous timescale in unit of Myr.
The vertical dashed lines mark the two values $t_{\rm v} = 1.7 \times 10^{5}$ yr and $t_{\rm v} = 1.1 \times 10^{4}$ yr, that give a radial inflation of $5\%$ (horizonatal dashed line). {\em Lower panel:}
Asynchronism as a function of viscous time in Myr, for J0651. The horizontal dashed lines mark the asynchronism expected for the two viscous timescales (vertical dahsed lines) for which the radial inflation of the secondary in J0651 is $5\%$ (see upper panel).}
\label{fig:doublepanel}
\end{figure}

The secondary's radius in J0651 has been measured precisely: from the eclipses of its lightcurve, Brown et al. (2011) derive $R_2 = 0.0353 \pm 0.0004 R_{\odot}$, noting
that this differs by $5\%$ from the $\approx 0.0337 R_{\sun}$ radius predicted by the best-matching ELM He WD model (Panei et al. 2007). Given the high
accuracy of the radius determination, this difference is significant at the $4 \sigma$ level. If taken at face value this discrepancy can be interpreted as due to tidal heating of the white dwarf, with two possible values of the viscous time ($\approx 170$ kyrs and $\approx 11$ kyr, cf. Fig. \ref{fig:doublepanel}, upper panel) reflecting the double role of viscosity in determining $W$ (see discussion in section 4.2, and Fig. \ref{fig:W-tv}). For a WD mass of $\simeq 0.25$ M$_{\odot}$ and radius $\approx 2.5 \times 10^9$ cm (Brown et al. 2011), these timescales correspond to an average\footnote{We adopt here the average WD density to derive this estimate. This is somewhat different from the average viscosity defined by, e.g. Press, Wiita \& Smarr (1975) and references therein.} $\mu \approx 9 \times 10^9$ g cm$^{-1}$ s$^{-1}$ and $\approx 1.5\times 10^{11}$ g cm$^{-1}$ s$^{-1}$, respectively.  This is orders of magnitude larger than the microscopic viscosity of the degenerate plasma\footnote{a dynamical viscosity of $\simeq 10^2$ g cm$^{-1}$ s$^{-1}$ would correspond to an average density 
$\lesssim 10^4$ g cm$^{-3}$.} (Kopal 1968, Durisen 1973), but somewhat smaller that the estimated viscosity of high-mass stars in X-ray binaries with circularized orbits (Sutantyo 1974). 
While fluid turbulence might play a role in WD interiors (Iben, Tutukov \& Fedorova 1998), it is also plausible that the viscosity associated to a properly oriented internal magnetic field of $\sim 10-100$  Gauss can suffice to reach this level of viscous dissipation (cfr. Sutantyo 1974, Smarr \& Blandford 1976).  

Although the viscous timescale is not uniquely determined by the amount of dissipation, this degeneracy is completely removed by a joint determination of the degree of 
synchronism of the white dwarf (Fig. \ref{fig:doublepanel}, lower panel). In the above interpretation the secondary's spin in J0651 would be expected to be $\sim$ 10 times slower than the orbital period for the longer value of t$_{\rm v}$, or to differ from it by only $\sim 20$ \%, in the alternative case. Viscous dissipation would also enhance the WD luminosity by $\lesssim 0.05$ L$_{\odot}$, which corresponds to a $\sim$ (30-50)\% perturbation of its intrinsic luminosity. Hence, along with a $\sim 5$\% increase of the outer radius, we would expect a $\sim$ (5-10)\% increase of the effective temperature. 

The above argument can be reversed to conclude that if the coefficient of dynamic viscosity were $\mu > 10^{12}$ g cm$^{-1}$ s$^{-1}$, as previously proposed in the literature, 
then the secondary's spin would be expected to be almost exactly synchronous, the amount of tidal dissipation would be much smaller than stated above and the inflation of the
white dwarf envelope essentially negligible. 
Therefore measuring the spin of the tidally perturbed secondary in J0651 will allow a straightforward check of these predictions, reaching unambiguous conclusions about the magnitude and nature of the internal viscosity of white dwarfs.

In Figure  \ref{fig:drsur}, we plot also the fractional size of the Roche lobe with respect to $R_2$: $(r_{\rm L} - R_2)/R_2$, where
$r_{\rm L} \approx 0.31 a$ (Eggelton 1983). We find that, due to the inflation of the outer envelope caused by tidal heating,  the secondary fills its Roche lobe ($dR_2 \approx r_{\rm L}$) when the binary has a period of $P \approx 9.7$ min 
and $P \approx 9$ min, for $t_{\rm v}\approx 1.7 \times 10^{5}$ yr and $\approx 1.1 \times 10^{4}$ yr respectively. This is well before the tidal radius ($P\approx 56 $ s) as defined in sec. \ref{sec:tidal_deformation}.

 %%%%%%%%%%%%%%%%%%%%%%%%%%%%%%%%%%%%%%%%%%%%%%%
%%%%%%%%%%%%%%%%%%%%%%%%%%%%%%%%%%%%%%%%%%%%%%%%%%%%%%%%%%%%%

 %%%%%%%%%%%%%%%%%%%%%%%%%%%%%%%%%%%%%%%%%%%%%%%
%%%%%%%%%%%%%%%%%%%%%%%%%%%%%%%%%%%%%%%%%%%%%%%%%%%%%%%%%%%%%

\section{Discussion on J0651 and Conclusions}
\label{sec:discussion}

We have thoroughly addressed the consequences of tidal interactions in detached white dwarf binaries, developing a simple and general formalism that includes the effects of viscosity and of GW-driven orbital evolution. The main physical implications of our scenario were illustrated quantitatively with a focus on short period systems, 
where tidal effects are most pronounced.

When applied to J0651, an ELM system with an orbital period of only $\sim$ 12.8 min, our analysis shows the full potential of tidal interactions to reveal the internal properties 
of white dwarfs. We have demonstrated that the large tidal deformation of the secondary star in J0651 can only be explained by including, in the tidal interaction energy, a term that depends on the star's ``deformability'', measured by the Love number ($\kappa_2$). This provides a direct observational test of the formulae derived here and in Paper I, hence also of the expression for the tidal radius (see $\S$ \ref{sec:tidal_deformation}). This approach will make it possible to probe the white dwarf internal structure when a sample of tidally distorted objects will be available, ranging in mass and orbital separations.

Further developing on the system's dynamics we have illustrated the role of the internal viscous torque, and of the GW-driven torque, in driving tidal coupling, hence the exchange of angular momentum between the orbit and the secondary star, and the double role of viscous dissipation in determining the synchronisation of the secondary's spin and the rate of internal heating. The latter 
causes the expansion of the non-degenerate stellar envelope, which ultimately determines the observed radius and luminosity of the white dwarf. 

In the binary system J0651, the eclipses provide a precise measurement of the radius of the secondary $R_2$ which we used in our application (see Tab.1). 
This value is $5\%$ larger than predicted by the best matching He WD model. 
We propose here that the discrepancy between the observed and predicted radius is factual, and can be due to tidal  inflation of the white dwarf outer layers. 
Based on this hypothesis we derive two possible values of the viscous time,  t$_{\rm v} \approx 170$ kyr or 11 kyr (corresponding to an average viscosity $\mu \approx 9 \times 10^9$ g cm$^{-1}$ s$^{-1}$, or  $\mu \approx 1.5 \times 10^{11}$ g cm$^{-1}$ s$^{-1}$). These viscosities may be due to a transverse magnetic field of $B \simeq 10-100$ Gauss in the white dwarf interior (Sutyanto 1974, Smarr \& Blandford 1976), or to tidally-driven turbulence in the WD interior (Press, Wiita \& Smarr 1975, Iben, Tutukov \& Fedorova 1998). We predict that these two candidate viscous timescales, should give synchronisation of $\alpha \approx 0.125$ and $\alpha \approx 0.8$, respectively.
Tentatively, we may exclude a middle range viscous time of a few $10^4$ yr ($\mu \lesssim 10^{11}$ g cm$^{-1}$ s$^{-1}$). This corresponds to the maximum achievable value of $W$ for J0651, and would give a rather large radial inflation of $\sim 10\%$ (see Fig. \ref{fig:drsur}).
If somewhat larger viscosities proposed in the literature were adopted, the degree of synchronism of the secondary in J0651 would be very close to unity. Accordingly, viscous dissipation would be negligible and would have no measurable effects on the white dwarf radius and/or effective temperature (see also discussion in section \ref{sec:inflation}).
%Remarkably, for such large viscosities the measured orbital shrinking of this system %would be mainly due to the tidal torque, with the emission of GW playing a secondary role %(see our discussion at the end of $\S$ \ref{sec:dynamics}.)

It then follows that a measurement of the asynchronism in this system would provide a crucial test for our model and, in particular, on the magnitude and nature of the internal viscosity of the secondary white dwarf. With a sample of short period detached WDs, with well measured masses and radii, there would be the intriguing possibility to determine whether the viscous timescale is unique -- hence the same viscous processes are effective in all WDs -- or there exist dependences on mass and chemical composition.

\section*{Acknowledgements}
SD was supported by an ERC Advanced Research Grant and by the Israeli Center for Excellence for High Energy Astrophysics.

\label{lastpage}
\end{document}